\documentclass[conference]{IEEEtran}
\IEEEoverridecommandlockouts

\usepackage[T1]{fontenc}

\usepackage{cite}
\usepackage{amsmath,amssymb,amsfonts}
\usepackage{graphicx}
\usepackage{textcomp}
\usepackage{xcolor}

\usepackage{acronym}
\usepackage{bbm}

\usepackage[]{subfig}

\usepackage{framed}

\usepackage{algorithm} 
\usepackage{algpseudocode}

\usepackage{hyperref}
\usepackage{url}
\usepackage[capitalize]{cleveref}

\crefformat{equation}{(#2#1#3)}
 \acrodef{AWGN}{additive white Gaussian noise}
\acrodef{ASK}{Amplitude Shift Keying}
\acrodef{iid}{independent and identically distributed}
\acrodef{SNR}{Signal-to-Noise Ratio}
\acrodef{MSE}{Mean-Squared Error}
\acrodef{MAE}{Mean Absolute Error}
\acrodef{SPRT}{Sequential Probability Ratio Test}
\acrodef{MMSE}{Minimum Mean-Squared Error}
\acrodef{MSPRT}{Matrix Sequential Probability Ratio Test}
\acrodef{BPSK}{Binary Phase Shift Keying}
\acrodef{pdf}{probability density function}
\acrodef{LP}{Linear Program} 

\DeclareMathOperator{\E}{\mathsf{E}} 		
\DeclareMathOperator{\Var}{\mathsf{Var}}

\newcommand{\Hyp}{\mathrm{H}} 	

\newcommand{\given}{\,|\,}
\newcommand{\bgiven}{\,\big|\,}

\newcommand{\param}{\theta}
\newcommand{\paramRV}{\Theta}

\newcommand{\dInt}{\mathrm{d}}	
\newcommand{\indd}[1]{\ensuremath{\mathbbm{1}{#1}}} 
\newcommand{\ind}[1]{\indd{\{#1\}} } 
\newcommand{\stat}[1]{\ensuremath{t_{#1}}} 
 
\newcommand{\transkernel}[1][n]{\ensuremath{\xi_{\stat{#1}}}}

\newcommand{\detConstr}[1][i]{\bar{\alpha}^{#1}}
\newcommand{\estConstr}[1][i]{\bar{\beta}^{#1}}

\newcommand{\SeqDataRV}{\ensuremath{\mathcal{X}_N}}

\newcommand{\stopR}{\ensuremath{\Psi} }

\newcommand{\dec}{\ensuremath{\delta}}

\newcommand{\stopAt}[1][]{\ensuremath{\Phi_{n#1}}}

\newcommand{\costDet}[1][n]{\ensuremath{\lambda}}
\newcommand{\costEst}[1][n]{\ensuremath{\mu}}
\newcommand{\costDetOpt}[1][n]{\ensuremath{\lambda^\star}}
\newcommand{\costEstOpt}[1][n]{\ensuremath{\mu^\star}}

\newcommand{\est}[1]{\ensuremath{\hat\param_{#1}}}

\newcommand{\errorDet}[2]{\ensuremath{\alpha_{#1}^{#2}}}
\newcommand{\errorEst}[2]{\ensuremath{\beta_{#1}^{#2}}}

\newcommand{\norm}[2]{\ensuremath{\mathcal{N}\left(#1,#2\right)}}
\newcommand{\Bern}{\mathrm{Bern}}

\newcommand{\stateSpaceHyp}[1][]{\ensuremath{E_\Hyp}}

\newlength{\imgWidthSingle}
\newlength{\imgWidthDouble}
\newcommand{\graph}{\mathcal{G}}
\newcommand{\graphFull}{(\nodeSet,\edgeSet)}
\newcommand{\nodeSet}{\mathcal{E}}
\newcommand{\edgeSet}{\mathcal{V}}

\newcommand{\Oneighborhood}[1][k]{\mathcal{N}_{#1}^{\text{o}}}
\newcommand{\Cneighborhood}[1][k]{\mathcal{N}_{#1}^{\text{c}}}

\newcommand{\OneighborhoodDef}[1][k]{\big\{l\in\nodeSet\given(#1,l)\in\edgeSet\bigr\}}
\newcommand{\CneighborhoodDef}[1][k]{\Oneighborhood[#1] \cup \{#1\}}

\newcommand{\SeqDataRVdistr}[1][k]{\SeqDataRV^{#1}}
\newcommand{\RVdistr}[2]{\ensuremath{X_{#1}^{#2}}}

\newcommand{\stopRdistr}[1][k]{\stopR^{#1}}
\newcommand{\decRdistr}[1][k]{\dec^{#1}}
\newcommand{\estdistr}[3]{{\hat\param}_{#1,#2}^{#3}}

\newcommand{\policyDistr}[1][k]{\pi^{#1}}
\newcommand{\policyDistrFull}[1][k]{\ensuremath{\{\stopR_n^{#1}, \dec_n^{#1}, \est{0,n}^{#1},\est{1,n}^{#1}\} _{0\leq n \leq N}}}

\newcommand{\statVec}[1]{\mathbf{t}_{#1}}

\newcommand{\statMod}[1]{\bar t_{#1}}
\newcommand{\inVec}[1]{\mathbf{x}_{#1}}
\newcommand{\combinedInno}[2]{\bar{x}_{#2}^{#1}}

\newcommand{\weightMtx}{\mathbf{W}}
\newcommand{\modWeightMtx}{\tilde{\mathbf{W}}}
\newcommand{\varMtx}[1]{\boldsymbol\Sigma_{\statVec{#1}}}

\newcommand{\transkernelDistr}[2]{\ensuremath{\xi_{\stat{#1}^{#2}}}}

\def\BibTeX{{\rm B\kern-.05em{\sc i\kern-.025em b}\kern-.08em
    T\kern-.1667em\lower.7ex\hbox{E}\kern-.125emX}}
\algnewcommand{\Inputs}[1]{
  \State \textbf{inputs:} #1
}
\algnewcommand{\Initialize}[1]{
  \State \textbf{initialize:} #1
}

\graphicspath{{./img/}}

\begin{document}

\title{Distributed Joint Detection and Estimation: A Sequential Approach
\thanks{The work of Dominik Reinhard was supported by the German Research Foundation (DFG) under grant number 390542458.}
\thanks{The work of Michael Fau\ss{} was supported by the German Research Foundation (DFG) under grant number 424522268.}
}

\author{\IEEEauthorblockN{Dominik Reinhard}
\IEEEauthorblockA{\textit{Signal Processing Group} \\
\textit{Technische Universit\"at Darmstadt}\\
64283 Darmstadt, Germany \\
reinhard@spg.tu-darmstadt.de}
\and
\IEEEauthorblockN{Michael Fau\ss{}}
\IEEEauthorblockA{\textit{Department of Electrical Engineering} \\
\textit{Princeton University}\\
Princeton, NJ 08544, USA \\
mfauss@princeton.edu}
\and
\IEEEauthorblockN{Abdelhak M. Zoubir}
\IEEEauthorblockA{\textit{Signal Processing Group} \\
\textit{Technische Universit\"at Darmstadt}\\
64283 Darmstadt, Germany \\
zoubir@spg.tu-darmstadt.de}
}

\maketitle

\begin{abstract}
We investigate the problem of jointly testing two hypotheses and estimating a random parameter based on data that is observed sequentially by sensors in a distributed network. 
In particular, we assume the data to be drawn from a Gaussian distribution, whose random mean is to be estimated. Forgoing the need for a fusion center, the processing is performed locally and the sensors interact with their neighbors following the consensus+innovations approach. We design the test at the individual sensors such that the performance measures, namely, error probabilities and mean-squared error, do not exceed pre-defined levels while the average sample number is minimized.  After converting the constrained problem to an unconstrained problem and the subsequent reduction to an optimal stopping problem, we solve the latter utilizing dynamic programming. The solution is shown to be characterized by a set of non-linear Bellman equations, parametrized by cost coefficients, which are then determined by linear programming as to fulfill the performance specifications. A numerical example validates the proposed theory.
\end{abstract}

\acresetall

\section{Introduction}
In many signal processing applications, detection and estimation occur in a coupled manner and both outcomes are of interest. That is, we want to perform a hypothesis test and, based on its outcome, estimate several parameters of the underlying model. Since solving the estimation and detection problem separately does not result in an overall optimal performance \cite{moustakides2012joint}, this problem has to be solved jointly. This line of work, called \emph{joint detection and estimation}, goes back to the 1960\cite{Middleton1968Simultaneous}. Later on, the problem was extended to  multiple hypotheses \cite{fredriksen1972simultaneous}. More recently, joint detection and estimation was applied to speech processing \cite{Momeni2015Joint}, change point detection\cite{boutoille2010hybrid}, communications \cite{wei2018simultaneous} and biomedical engineering\cite{chaari2013fast}.

Sequential analysis is a field of research pioneered by Abraham Wald in the 1940 with the \ac{SPRT}\cite{wald1947sequential}. In that framework, the inference should be performed as fast as possible, while fulfilling pre-defined constraints on the quality of the outcome. Sequential inference is an area of ongoing research. Sequential methods can reduce the number of samples up to $50\,\%$ on average. Thus, such approaches are preferable to conventional ones, especially in low-power or time critical applications. An overview on sequential detection and estimation is given in \cite{tartakovsky2014sequential} and \cite{ghosh2011sequential}, respectively.

Combining the ideas of joint detection and estimation with those of sequential analysis leads to a framework in which as few samples as possible are used on average while the true hypothesis and random parameters are inferred simultaneously. This topic was investigated, for example, in \cite{yilmaz2015sequential,yilmaz2016Sequential}. In those works, the aim is to minimize the number of used samples under the constraint that a combined cost function, which incorporates detection and estimation errors, is kept below a certain level. We investigated the problem of joint detection and estimation under distributional uncertainties in \cite{reinhard2016}. In \cite{reinhard2018bayesian}, we proposed a Bayesian framework in which the \emph{average} number of samples is minimized under the constraints that the detection and estimation errors are kept below certain levels. That framework was later applied to joint signal detection and \ac{SNR} estimation\cite{reinhard2019jointSNR}. In \cite{reinhard2020jointSymbolDecoding}, we proposed an approach for sequential joint detection and estimation under multiple hypotheses and applied it to joint symbol decoding and noise power estimation.

In many modern applications, a network of multiple sensors is used to gather data. There exist three types of sensor networks: centralized, decentralized and distributed. In a centralized network, the entire processing is done at a single processing unit, the fusion center. In decentralized sensor networks, some of the processing is done at the sensors, but the final inference is still left to the fusion center, such as in \cite{goelz2019}. In a distributed sensor network, the entire processing is performed at the nodes through local interactions in the neighborhood, such as in \cite{kar2013Consensus,leonard2018Robust,li2016,sahu2015distributed,xiao2004fast}. Distributed sensor networks have the advantage that there is no single point of failure, making them robust against link and sensor faults.

In this work, we investigate the problem of sequential joint detection and estimation in a distributed sensor network. The sensors cooperate via the \emph{Consensus+Innovations} approach \cite{kar2013Consensus}. The optimal policy at a particular node is designed by exploiting the theoretical results in \cite{reinhard2018bayesian}. To the best of our knowledge, joint detection and estimation in distributed sensor networks has not been treated in the literature yet, neither in a sequential nor in a fixed sample size setup.

The remainder of the work is structured as follows. In \cref{sec:problemFormulation}, a detailed problem formulation is given and the fundamentals of sequential joint detection and estimation are briefly revised. The representation of the data and the local interaction between the nodes is described in \cref{sec:informExchange}. In \cref{sec:solution}, the solution of the design problem is presented, which is validated by a numerical example in \cref{sec:results}. Concluding remarks are given in \cref{sec:conclusion}.
 \section{Problem Formulation}\label{sec:problemFormulation}
Consider a network of $K$ sensors, which can be modeled by a simple, connected and undirected graph $\graph=\graphFull$, where the sets of nodes and edges are denoted by $\nodeSet$ and $\edgeSet$, respectively. Let $\Oneighborhood=\OneighborhoodDef$ and $\Cneighborhood=\CneighborhoodDef$ denote the open and closed neighborhood of node $k$, respectively. Each sensor $k\in\{1,\ldots,K\}$ observes a sequence of random variables $\SeqDataRVdistr=(\RVdistr{1}{k},\ldots,\RVdistr{N}{k})$, which can be generated under two different hypotheses $\Hyp_0$ or $\Hyp_1$.
Under both hypotheses, the data conditioned on the mean is Gaussian distributed with variance $\sigma^2$. Moreover, the mean is itself a random variable, which follows a known distribution under both hypotheses. The priors have a disjoint support and the occurrence of the hypotheses is a random variable with known probability $p(\Hyp_i)$, $i\in\{0,1\}$. Mathematically, the two hypotheses can be written as:
\begin{align*}
 \Hyp_0:& \quad \SeqDataRVdistr\given\param_0\sim\norm{\param_0}{\sigma^2}\,,\;\paramRV_0\sim p(\param_0) \\
 \Hyp_1:& \quad \SeqDataRVdistr\given\param_1\sim\norm{\param_1}{\sigma^2}\,,\;\paramRV_1\sim p(\param_1)
\end{align*}
In this work, each sensor should jointly infer the true hypothesis as well as the underlying parameter in a sequential manner. That is, each sensor observes the sequence $\SeqDataRVdistr$ sample by sample and makes a decision as soon as the confidence about the true hypothesis and the parameter is high enough. To perform the inference task, each sensor uses its own data as well as information from its neighborhood.

The following assumptions are made throughout the paper: First, the random variables $\RVdistr{n}{k}\given\Hyp_i,\param_i$ are \ac{iid} for all $n=1,\ldots,N$ and all $k=1,\ldots,K$. Second, all sensors share the same realization of the random parameter and the hypothesis, which stay  constant during the observation period.

\subsection{Fundamentals of Sequential Joint Detection and Estimation}
At each time instant $n$ and at every sensor $k$ a stopping rule $\stopRdistr_n\in\{0,1\}$ is evaluated to decide whether sensor $k$ is certain enough about the true hypothesis and the unknown parameter. Once the stopping rule of sensor $k$ takes on the value $1$, this sensor jointly infers the hypothesis and the true parameter. More precisely, the decision rule $\decRdistr_n\in\{0,1\}$ is evaluated to obtain the hypothesis; to obtain an estimate at node $k$, the estimator $\estdistr{i}{n}{k}$ is applied, where $i$ denotes the value of $\decRdistr_n$. The collection of stopping rule, decision rule and the two estimators is referred to as policy. At node $k$, the policy is given by $\policyDistr = \policyDistrFull$.
The time instant at which sensor $k$ stops, i.e., the stopping time, is defined as
\begin{align*}
 \tau^k = \min\Bigl\{n\geq1: \stopRdistr_n=1\Bigr\}\,.
\end{align*}
To measure the individual performance of node $k$, we use the average sample number $\E\bigl[\tau^k\big]$ along the error probabilities and the \ac{MSE}, which are given by
\begin{align*}
  \errorDet{i}{k} & = \E\Bigl[\ind{\decRdistr_{\tau^k}\neq i}\bgiven\Hyp_i\Bigr]\,,\quad i\in\{0,1\}\,,\\
  \errorEst{i}{k} & = \E\Bigl[\ind{\decRdistr_{\tau^k}=i}\bigl(\param - \estdistr{i}{\tau^k}{k}\bigr)^2\bgiven\Hyp_i\Bigr]\,,\quad i\in\{0,1\}\,.
\end{align*}
In the previous equations, $\ind{\cdot}$ denotes the indicator function.

\subsection{Test Design as an Optimization Problem}
We only consider truncated sequential schemes in this work, i.e., scheme which use at most $N$ samples, which implies that $\stopRdistr_N=1$ for all $k=1,\ldots,K$.

The aim is to find a sequential scheme which uses on average as few samples as possible, while the type I and type II error probabilities as well as the \acp{MSE} under both hypothesis are limited to pre-defined levels. Since designing a \emph{global} optimal policy is infeasible, we formulate the design problem locally as in, e.g., \cite{leonard2018Robust,sahu2015distributed}. That is, for each node $k$, we want to design a scheme which uses  a minimum number of samples on average, while the detection and estimation errors are kept below pre-defined levels.
However, although the design problem is formulated at the sensor level rather than on the network level, there exist a coupling through the neighborhood communication. 

Mathematically, the design problem for node $k$ can be formulated as the following optimization problem
\begin{align}\begin{split}\label{eq:constrProblem}
 & \min_{\policyDistr}\;\E\bigl[\tau^k\bigr], \quad \stopRdistr_N=1\\
 \text{s.t.}\quad & \errorDet{i}{k} \leq \detConstr[]_i\,,\quad i=0,1\,, \\
 & \errorEst{i}{k} \leq \estConstr[]_i\,,\quad i=0,1\,,
 \end{split}
\end{align}
where $\detConstr[]_i\in(0,1)$ and $\estConstr[]_i\in(0,\infty)$ are the maximally tolerated detection and estimation errors, respectively.

Instead of solving \cref{eq:constrProblem} directly, we first consider the following auxiliary problem
\vspace*{-5pt}
\begin{align}\label{eq:unconstrProblem}
 \min_{\policyDistr}\;\biggl\{\E\bigl[\tau^k\bigr] + \sum_{i=0}^1 p(\Hyp_i)\Bigl(\costDet_i^k\errorDet{i}{k} + \costEst_i^k\errorEst{i}{k}\Bigr)\biggr\}\,,
\end{align}
where $\costDet_i^k$ and $\costEst_i^k$, $k=1,\ldots,K$, $i=0,1$, are some non-negative and finite cost coefficients. The choice of the cost coefficients is discussed later in the paper.

 \section{Data Representation and Information Exchange}\label{sec:informExchange}
Since each sensor observes the data sequentially, the sample space grows with every new sample. For the reason of tractability, a low-dimensional representation of the data has to be found. Here, the sample mean of the data is used as a representation since it contains all information relating the samples and the random mean for a Gaussian likelihood \cite{reinhard2018bayesian}.
Although \cite{reinhard2018bayesian} uses the sample mean directly as state variable for the test, this work is in a distributed setup and hence, the state of a sensor comprises not only information from its measurements, but also from the state of its neighbors. More precisely, we resort to the \emph{consensus+innovations} approach \cite{kar2013Consensus}, which was already applied to distributed sequential detection \cite{li2016,leonard2018Robust}.

Let $\stat{n}^k$ be the state of sensor $k$ at time $n$, with a initial state $\stat{0}^k=0$ for all $k=1,\ldots,K$. The state update at sensor $k$ is then given by
\vspace*{-5pt}
\begin{align}\label{eq:CI_scalar}
 \stat{n}^k = \sum_{l\in\Cneighborhood} \biggl( \frac{n-1}{n} w_{kl}\stat{n-1}^{l}  + \frac{1}{n}w_{kl}x_{n}^l\biggr)\,,
\end{align}
where $w_{kl}$ are some appropriate weights, the choice of which will be discussed later. In \cref{eq:CI_scalar}, the weighted sum of the states is called the \emph{consensus} term and the weighted sum of the observations is the \emph{innovations} term.
By stacking the states and the observations in vectors, i.e., $\statVec{n} = [\stat{n}^1,\ldots,\stat{n}^K]^\top$ and $\inVec{n} = [x_{n}^1,\ldots,x_{n}^K]^\top$, and collecting the weights $w_{kl}$ in a matrix $\weightMtx$, the update of all states can be rewritten as
\begin{align*}
 \statVec{n} & = \boldsymbol W \biggl( \frac{n-1}{n} \statVec{n-1} + \frac{1}{n} \inVec{n}\biggr) = \frac{1}{n}\sum_{i=1}^n \weightMtx^{n-i+1} \inVec{i}\,.
\end{align*}
\subsection{Choice of Weights}
From a theoretical point of view, the weighting matrix $\weightMtx$ can be arbitrary as long as it is right stochastic, i.e.,
\begin{align*}
 \sum_{i\in\Cneighborhood} w_{kl} = 1\,,\; \forall k, \quad \text{and} \quad w_{kl} \geq 0\,,\; \forall k,l\,.
\end{align*}
In \cite{li2016}, the authors assumed that the matrix $\weightMtx$ to be non-negative, symmetric, irreducible and stochastic. It is designed as $\weightMtx = \mathbf{1} - c\mathbf{L}$,
where $\mathbf{1}$ denotes the identity matrix, $\mathbf{L}$ is the graph Laplacian matrix and $c$ some suitably chosen constant. This approach was first introduced in \cite{xiao2004fast}. However, this design procedure often places negative weights on the main diagonal \cite{xiao2004fast, leonard2018Robust}.

In this work, we use equal weights for the closed neighborhood of a node, i.e.,
\vspace*{-2pt}
\begin{align*}
 w_{kl} = \begin{cases}
            \frac{1}{\lvert\Cneighborhood\rvert}, & l\in\Cneighborhood\\
            0 & \text{otherwise}
          \end{cases}
\end{align*}
This more intuitive weight design was already used in the context of distributed sequential detection \cite{leonard2018Robust}. 
\subsection{Statistical Properties of the State Variable}\label{sec:statPropState}
Before the solution methodology is presented, a statistical characterization of the random state variable is presented. Since the random variable $\stat{n}^k\given\param_i,\Hyp_i$ is a linear combination of Gaussian random variables, it is again a Gaussian random variable. Hence, it is sufficient to calculate its mean and its variance.
Following the arguments in \cite[Section IV-A]{leonard2018Robust}, it can be shown that the mean and the variance are given by
\begin{align*}
 \E[\stat{n}^k \given \Hyp_i, \param_i] & = \param_i\,, \\
 \Var[\stat{n}^k \given \Hyp_i, \param_i] & = \frac{\sigma^2}{n^2} \sum_{i=1}^n \mathbf{e}^\top_k\weightMtx^i\bigl(\weightMtx^i\bigr)^\top \mathbf{e}_k := \sigma^2_{\stat{n}^k}\,,
\end{align*}
where $\mathbf{e}_k$ is the $k$th column of the identity matrix $\mathbf{1}$.
Hence, we can state that $\stat{n}^k\given\Hyp_i,\param_i\sim\norm{\param_i}{\sigma^2_{\stat{n}^k}}$.
 \section{Solution Methodology}\label{sec:solution}
In order to design the tests, we rely on our framework proposed in \cite{reinhard2018bayesian}: First, \cref{eq:unconstrProblem} has to be reduced to an optimal stopping problem, which is then solved by means of dynamic programming. The coefficients, which parametrize the solution of the optimal stopping problem are then  obtained via linear programming, such that the solutions of \cref{eq:constrProblem} and \cref{eq:unconstrProblem} coincide.

\subsection{Reduction to an Optimal Stopping Problem}
To end up with an optimal stopping problem, the unconstrained problem in \cref{eq:unconstrProblem} first has to be minimized with respect to the decision rule and then with respect to the estimators\cite{reinhard2018bayesian}. With the optimal decision rule and the optimal estimators, the design problem reduces to the optimal stopping problem
\begin{align}\label{eq:optStoppingProbl}
\min_{\stopR^k} \sum_{n=0}^N\E\bigl[\stopAt^k(n+g^k(\stat{n}^k))\bigr]\,, 
\end{align}
with the short hand notation $\stopAt^k = \stopR^k_n\prod_{i=0}^{n-1}(1-\stopR^k_i)$ and the instantaneous cost
\begin{align}\label{eq:defG}
 g^k(\stat{n}^k) = \min\bigl\{D_{0,n}^k(\stat{n}^k)\,,\,D_{1,n}^k(\stat{n}^k)\bigr\}\,.
\end{align}
The instantaneous cost $g^k(\stat{n}^k)$ for stopping at time $n$ is the minimum of $D_{0,n}^k(\stat{n}^k)$ and $D_{1,n}^k(\stat{n}^k)$, which are the costs for stopping at time $n$ and deciding in favor of $\Hyp_0$ and $\Hyp_1$, respectively. The cost for stopping and deciding in favor of $\Hyp_i$ is given by
\begin{align*}
 D_{i,n}^k(\stat{n}^k) = \costDet_{1-i}^kp(\Hyp_{1-i}\given\stat{n}^k) + \costEst_{i}^kp(\Hyp_{i}\given\stat{n}^k)\Var[\paramRV_i\given\Hyp_i,\stat{n}^k]\,.
\end{align*}
These costs consist of two parts. The first part penalizes wrong decisions, whereas the second part penalizes inaccurate estimates.

\subsection{Characterization of the Cost Function}
For fixed coefficients $\costDet_{i}^k$, $\costEst_{i}^k$, $i\in\{0,1\}$, the solution of the optimal stopping problem in \cref{eq:optStoppingProbl} can be characterized by a set of non-linear Bellman equations, see, e.g., \cite{fauss2015linear, reinhard2018bayesian}, 
\begin{align*}
 \rho_n^k(\stat{n}^k) & = \min\Big\{g^k(\stat{n}^k)\,,\, d_n(\stat{n}^k)\Big\}\,, n<N\,,\\
 \rho_N^k(\stat{N}^k) & = g^k(\stat{N}^k)\,,
\end{align*}
with the cost for stopping as defined in \cref{eq:defG} and the cost for continuing the test
\begin{align}\label{eq:defD}
 d_n^k(\stat{n}^k) = 1 + \E[ \rho_{n+1}^k(\stat{n+1}^k) \given \stat{n}^k]\,.
\end{align}

\begin{table}[t!]
\vspace{0.2in}
  \caption{Sampling from the posterior predictive.}
  \label{alg:samplPostPred}
  \vspace{-0.1in}
\begin{framed}
\begin{algorithmic}[1]
    \For{$l=1,\ldots,N_\text{samp}$}
      \State sample $\Hyp^{(l)} \sim \Bern(r)$, with $r=p(\Hyp_1\given\stat{n}^k)$
      \State sample $\param^{(l)} \sim p(\param\given\Hyp^{(l)},\stat{n}^k)$
      \State sample $\statMod{n+1}^{k,(l)} \sim \norm{(1-w_{kk})\param^{(l)}}{\sigma^2_{\statMod{n+1}^k}}$
      \State sample $\combinedInno{k,(l)}{n+1} \sim \norm{\param^{(l)}}{\sigma^2_{\combinedInno{k}{n+1}}}$
    \EndFor
  \end{algorithmic}
 
\end{framed}

\end{table}

By defining
\begin{align*}
 \statMod{n+1}^k & = \sum_{l\in\Oneighborhood} w_{kl}\stat{n}^{l} \quad \text{and} \quad \combinedInno{k}{n+1} = \sum_{l\in\Cneighborhood} w_{kl}x_{n+1}^l\,,
\end{align*}
the state transition in \cref{eq:CI_scalar} can be rewritten as:
\begin{align}
 \stat{n+1}^k 
	      & = \frac{1}{n+1}\Bigl(\!nw_{kk} \stat{n}^k + n\statMod{n+1}^k + \combinedInno{k}{n+1}\Bigr) \nonumber := \transkernelDistr{n}{k}(\combinedInno{k}{n+1},\statMod{n+1}^k)
\end{align}
Hence, the cost for continuing is calculated as
\begin{align}\label{eq:calcD}
 d_n^k(\stat{n}^k) = 1 + \iint & \rho_{n+1}^k(\transkernel(\combinedInno{k}{n+1},\statMod{n+1}^k))\\
			      & \times p(\combinedInno{k}{n+1},\statMod{n+1}^k\given \stat{n}^k)\dInt\combinedInno{k}{n+1}\dInt\statMod{n+1}^k \,.
\end{align}
A derivation of the posterior predictive is laid down in \cref{app:postPred}. The integral in \cref{eq:calcD} can either be solved numerically or approximated by Monte Carlo integration. 
That is, the cost for continuing is approximated by
\begin{align*}
 d_n^{k}(\stat{n}^k) \approx 1 + \frac{1}{N_\text{samp}} \sum_{l=1}^{N_\text{samp}}\rho_{n+1}^k(\transkernel(\combinedInno{k,(l)}{n+1},\statMod{n+1}^{k,(l)}))\,,
\end{align*}
where the sequence $\Bigl\{\combinedInno{k,(l)}{n+1},\statMod{n+1}^{k,(l)}\Big\}_{l=1}^{N_\text{samp}}$ is obtained by sampling from the posterior predictive $p(\combinedInno{k}{n+1},\statMod{n+1}^k\given \stat{n}^k)$ as summarized in \cref{alg:samplPostPred}. In this algorithm, $\Bern(r)$ is the Bernoulli distribution with success rate $r$.

The policy of the optimal sequential scheme at node $k$, which is induced by $\rho^k_n$, can be summarized as:
\begin{align}\label{eq:optPolicy}
 \begin{split}
  \dec_{n}^k(\stat{n}^k) & = \ind{D_{0,n}^k(\stat{n}^k) > D_{1,n}^k(\stat{n}^k)} \\
  \stopR_{n}^k(\stat{n}^k) & = \ind{\rho_n^k(\stat{n}^k) = g^k(\stat{n}^k) }\\
  \estdistr{i}{n}{k} & = \E[\paramRV_i \given \Hyp_i, \stat{n}^k]\,, \quad i=0,1
 \end{split}
\end{align}
Since the functions $g^k(\stat{n}^k)$ and $\rho^k_n(\stat{n}^k)$ are parametrized by the coefficients $\costDet_i^k$, $\costEst_i^k$, $i=0,1$, the policy also depends on these coefficients. Hence, their choice is crucial for the performance.

\subsection{Choice of the Cost Coefficients}
For choosing the cost coefficients such that the resulting test also solves \cref{eq:constrProblem}, a strong connection between the cost function and the performance measures \cite[Theorem 4.2]{reinhard2018bayesian} \cite[Theorem 3.2]{fauss2015linear} is exploited. As outlined in \cite{fauss2015linear,reinhard2018bayesian}, the final design problem can be cast into a linear program, which can be efficiently solved by various off-the-shelf solvers. Substituting $\costDet^k=(\costDet_0^k,\costDet_1^k)$ and $\costEst^k=(\costEst_0^k,\costEst_1^k)$ gives the final optimization problem for the test at sensor $k$:
\begin{align}\label{eq:LP}
 \max_{\costDet^k\geq0,\costEst^k\geq0,\rho_n^k}
 \, & \biggl\{\rho_0^k(\stat{0}^k) - \sum_{i=0}^1p(\Hyp_i)(\costDet_i^k\detConstr + \costEst_i^k\estConstr)\biggr\}\\
 \text{s.t.} \quad
 &\rho_n \leq D^k_{i,n}\;,\; i\in\{0,1\}\,,\; n=0,\ldots,N\,, \nonumber\\
 &\rho_n \leq 1+\E[ \rho_{n+1}^k(\stat{n+1}^k) \given \stat{n}^k]\,,\; n<N\,,\nonumber
\end{align}
The resulting test fulfills the constraints with equality when the corresponding cost coefficients are positive. However, when a coefficient becomes zero, the corresponding constraint is still fulfilled implicitly, but not with equality. See \cite[Appendix F]{reinhard2018bayesian} for a detailed discussion. \section{Numerical Results}\label{sec:results}
\begin{figure}[t!]
 \begin{center}
  \includegraphics[scale=0.93]{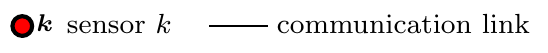}\\
  \includegraphics[scale=0.93]{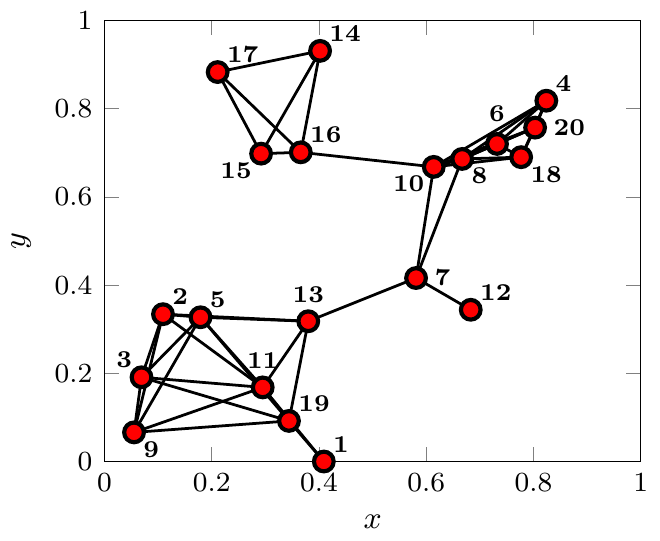}
  \caption{Network with $K=20$ agents and a communication radius of $d_\text{com}=0.3$}
  \label{fig:network}
 \end{center}
\end{figure}

We consider a network of $K=20$ sensors with $x$- and $y$-coordinates uniformly sampled on the interval $[0,1]$. The neighborhood of a node is defined as all nodes which are inside a communication radius of $d_\text{com}=0.3$. The network generation is repeated until the graph is connected. The generated network is depicted in \cref{fig:network}.
In order to validate the proposed approach, we chose the following example setup,
\begin{align*}
 \Hyp_0:& \quad \SeqDataRVdistr\given\param_0\sim\norm{\param_0}{4^2}\,,\;\paramRV_0\sim\norm{-2}{0.5^2}\,, \\
 \Hyp_1:& \quad \SeqDataRVdistr\given\param_1\sim\norm{\param_1}{4^2}\,,\;\paramRV_1\sim \norm{2}{0.5^2}\,,
\end{align*}
where both hypotheses have equal prior probabilities.
Although it was mentioned in \cref{sec:problemFormulation} that the priors $p(\Hyp_0)$ and $p(\Hyp_1)$ must have a disjoint support, this is not the case for Gaussian priors. Nonetheless, with this particular choice of hyperparameters, the support can be assumed to be almost disjoint. Moreover, the chosen prior acts as a conjugate prior, which allows us to derive a closed-form expression of the posterior distribution and to easily sample from it.

Each sensor should jointly infer the underlying hypothesis and parameter, while the detection and estimation errors are limited to $\detConstr[0]=\detConstr[1]=10^{-3}$ and $\estConstr[0] = \estConstr[1]=0.1$, respectively, and at most $N=50$ samples should be used.

To design the tests, the linear program in \cref{eq:LP} is solved using the gurobi optimizer\cite{gurobi}, which is called via the Matlab cvx interface\cite{cvx,gb08}. To solve the problem numerically, the state is discretized on the interval $[-9,9]$ with $1900$ equally spaced points.
The expectation in \cref{eq:defD} is approximated via Monte Carlo integration with $5\cdot10^4$ samples, which are generated according to the algorithm in \cref{alg:samplPostPred}. Note that, since we chose a conjugate prior, the posterior distribution in line $3$ is Gaussian distributed, i.e., $p(\theta\given\Hyp^{(l)},\stat{n}^k)=\mathcal{N}\Bigl(\param_\text{post},\sigma^2_\text{post}\Bigr)$ with $\param_\text{post}=\sigma^2_\text{post}\Bigl(\frac{\pm2}{0.5^2} + \frac{\stat{n}^k}{4^2}\Bigr)$ and $\sigma^2_\text{post}=\Bigl(0.5^{-2} + \sigma_{\stat{n}^k}^{-2}\Bigr)^{-1}$ and the sign of $\pm2$ depends on the value of $\Hyp^{(l)}$. Hence, we can easily sample from the posterior distribution.
\begin{figure}[t!]
 \centering
 \includegraphics[scale=0.9]{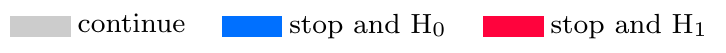}\\[.2em]
 \includegraphics[scale=0.9]{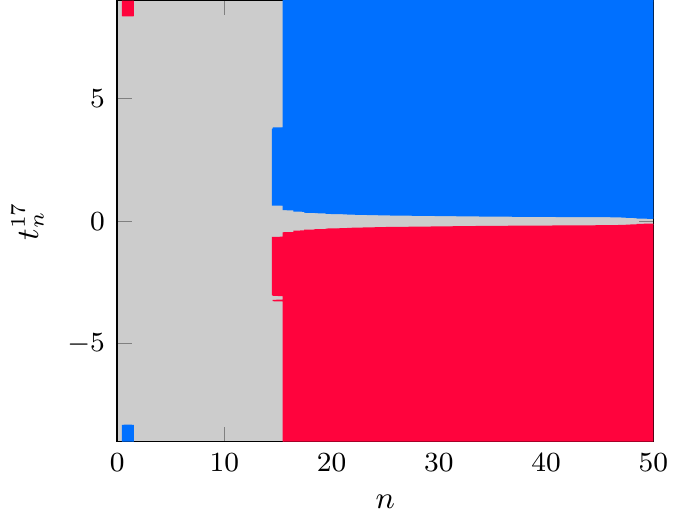}
 \caption{Policy of node $17$.}
 \label{fig:policy17}
\end{figure}

To illustrate the behavior of a designed test, the policy of node $17$ is depicted in \cref{fig:policy17}. The policy comprises three different regions, one for continuing the test (gray color) and one for stopping the test and deciding in favor of $\Hyp_0$ (blue color) and $\Hyp_1$ (red color), respectively. For $n\leq14$, the cost for making an inaccurate estimate dominates so that the procedure does not stop in this region. Subsequently, the test continues sampling only for small values of $\lvert\stat{n}^{17}\rvert$ since the uncertainty about the true hypothesis is high. Note that there exist two regions in which the tests stops for $n=1$, which are artifacts caused by numerical inaccuracies, e.g., too few samples for estimating the costs for continuing.

The designed tests are then validated with $10^6$ Monte Carlo runs.
The validation is two-fold. First, the performance on node-level, i.e., the performance of all individual nodes, is considered as it is the design goal in \cref{eq:constrProblem}.
Second, the performance is averaged over the network as in \cite{sahu2015distributed,leonard2018Robust}. That is, for each Monte Carlo run a random node is selected to evaluate the performance.

\begin{figure*}[t!]
  \centering
  \includegraphics{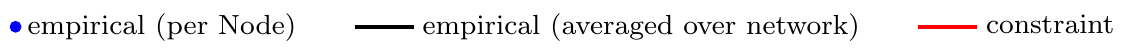}\\[-1em]
  \subfloat[Type I error.\label{fig:resDet0}]{\includegraphics{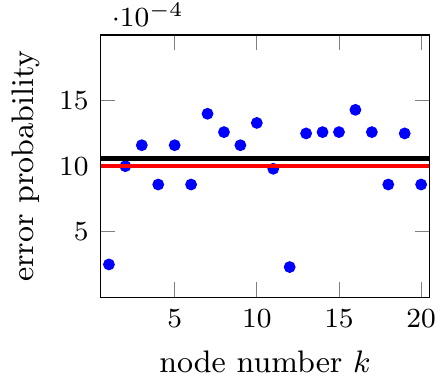}}
  \hfill
  \subfloat[Type II error.\label{fig:resDet1}]{\includegraphics{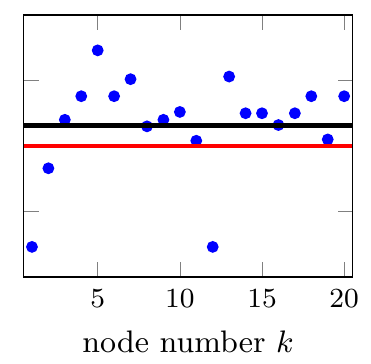}}
  \hfill
  \subfloat[\ac{MSE} under $\Hyp_0$.\label{fig:resEst0}]{\includegraphics{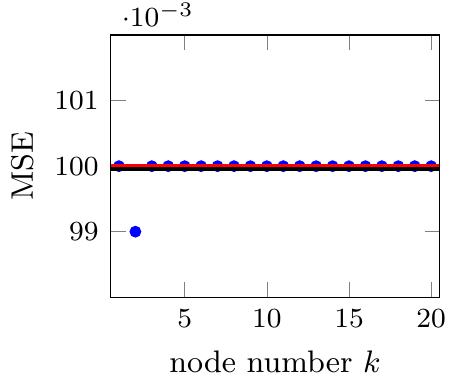}}
  \hfill
  \subfloat[\ac{MSE} under $\Hyp_1$.\label{fig:resEst1}]{\includegraphics{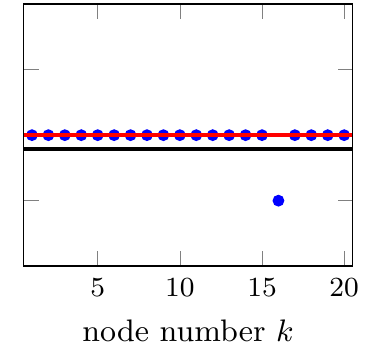}}
  \caption{Simulation results of the Monte Carlo simulation with $10^6$ runs. The empirical values of a single node are represented by blue circles, the averages over the network by black lines and the constraints by red lines.}
  \label{fig:simResults}
 \end{figure*}
 
The detection and estimation errors are depicted in \cref{fig:simResults}, where the detection/estimation errors of a single node are indicated by blue circles, the averaged errors over the network are indicated by black lines and the constraints are indicated by red lines. From \cref{fig:resEst0,fig:resEst1}, it can be seen that averaged \acp{MSE} as well as the \acp{MSE} of the individual nodes are close to the targeted values. The detection errors, on the other hand, have, under both hypotheses, a higher fluctuation over the nodes than the estimation errors. However, the individual as well as the averaged empirical error probabilities are reasonable close to the constraint. There are several reasons which can cause these small deviations. First, numerical inaccuracies arise during the design process. Especially the quality of approximation of the cost for continuing via Monte Carlo integration depends on the number of samples. Using $5\cdot10^4$ samples to approximate the integral might not be sufficient. Second, the assumption that the priors have a disjoint support is violated. Last, the number of Monte Carlo runs might be too small to represent the statistical properties of the observed data and its flow through the network.
 \section{Conclusion}\label{sec:conclusion}
We have investigated the problem of sequential joint detection and estimation in a distributed fashion for Gaussian distributed data with a random mean. The local interaction of the individual sensors follow a consensus+innovations approach.
The test at each node has been designed such that it is of minimum expected run-length and meets pre-defined levels of error probabilities and \acp{MSE}.
The proposed theory has been validated with a numerical example.  \appendix
\crefalias{section}{appsec}
\section{Derivation of the Posterior Predictive}\label{app:postPred}
The joint posterior predictive can be calculated via
\begin{align*}
  p(\statMod{n+1}^k,\combinedInno{k}{n+1}\given\stat{n}^k) = & \sum_{i=0}^1 p(\statMod{n+1}^k,\combinedInno{k}{n+1}\given\Hyp_i,\stat{n}^k)p(\Hyp_i\given\stat{n}^k)\,.
\end{align*}
Under hypothesis $\Hyp_i$, the posterior predictive is given by
 \begin{align*}
 p(\statMod{n+1}^k,\combinedInno{k}{n+1}\given& \Hyp_i,\stat{n}^k) \\
 & = \int p(\statMod{n+1}^l,\combinedInno{k}{n+1}\given\Hyp_i,\param_i) p(\param_i\given\Hyp_i,\stat{n}^k)\dInt\param_i\,.
\end{align*}
Since $\statMod{n+1}^k$ and $\combinedInno{k}{n+1}$ are conditionally independent given $\Hyp_i$ and $\param_i$, the joint likelihood factorizes as
\begin{align*}
  p(\statMod{n+1}^k,\combinedInno{k}{n+1}\given\Hyp_i,\param_i) = p(\statMod{n+1}^k\given\Hyp_i,\param_i) p(\combinedInno{k}{n+1}\given\Hyp_i,\param_i)\,.
\end{align*}
With the same line of arguments as in \cref{sec:statPropState}, one can show that
\vspace*{-5pt}
\begin{align*}
 \combinedInno{k}{n+1}\given\Hyp_i,\param_i \sim \norm{\param_i}{\sigma^2_{\combinedInno{k}{n+1}}}\,,
\end{align*}
where the variance is given by
\begin{align*}
 \sigma^2_{\combinedInno{k}{n+1}} := \Var[\combinedInno{k}{n+1}\given\Hyp_i,\param_i] = \sigma^2\mathbf{e}^\top_k\weightMtx\weightMtx^\top \mathbf{e}_k\,.
\end{align*}

In order to derive the mean and the variance of $\statMod{n+1}^k\given\Hyp_i,\param_i$, a modified weight matrix $\modWeightMtx$, whose diagonal elements are zeros, is introduced. The elements $\tilde w_{kl}$ of $\modWeightMtx$ are given by:
\begin{align*}
  \tilde w_{kl} = \begin{cases}
            w_{kl}\,,& k\neq l\\
            0\,, & k=l
          \end{cases} 
\end{align*}
Then, the combined states of the open neighborhood can be written as
\begin{align*}
 \statMod{n+1}^k = \sum_{l\in\Oneighborhood} w_{kl}\stat{n}^{l} = \mathbf{e}^\top \modWeightMtx \statVec{n}\,.
\end{align*}
It can be easily seen that the conditional mean is given by
\begin{align*}
 \E[\statMod{n+1}^k\given\Hyp_i,\param_i] =  (1-w_{kk})\param_i := \tilde\param_i^k\,.
\end{align*}
The conditional variance can be now be calculated as
\begin{align}\label{eq:varStatMod}
\begin{split}
 & \Var\Bigl[\statMod{n+1}^k\given\Hyp_i,\param_i\Bigr] \\
 \quad & = \mathbf{e}^\top \modWeightMtx \E\Bigl[\statVec{n}\statVec{n}^\top\given\Hyp_i,\param_i\Bigr] \modWeightMtx^\top\mathbf{e}_k - \bigl(\tilde\param_i^k\bigr)^2 \,.
 \end{split}
\end{align}
Similarly to \cite[Section IV-A]{leonard2018Robust}, it can be shown that:
\begin{align} \label{eq:EstatSq}
 \E[\statVec{n}\statVec{n}^\top \given\Hyp_i,\param_i] & = \frac{\sigma^2}{n^2}\sum_{i=1}^n \weightMtx^{i}\biggl(\weightMtx^{i}\biggr)^\top  + \param_i^2 \nonumber \\
 & = \varMtx{n} + \param_i^2
\end{align}
Inserting \cref{eq:EstatSq} into \cref{eq:varStatMod}, finally gives
\begin{align}\label{eq:varStatModFinal}
 \Var\Bigl[\statMod{n+1}^k\given\Hyp_i,\param_i\Bigr] = \mathbf{e}^\top \modWeightMtx \varMtx{n} \modWeightMtx^\top\mathbf{e}_k := \sigma^2_{\statMod{n+1}^k}\,.
\end{align}
Therefore, the likelihood of $\statMod{n+1}^k$ is given by
\begin{align*}
 p(\statMod{n+1}^k\given\Hyp_i,\param_i) = \norm{\tilde\param_i^k}{\sigma^2_{\statMod{n+1}^k}}\,.
\end{align*}
Hence, the joint likelihood under $\Hyp_i$ becomes
\begin{align*}
  p(\statMod{n+1}^k,\combinedInno{k}{n+1}\given\Hyp_i,\param_i) & =  \norm{\tilde\param_i^k}{\sigma^2_{\statMod{n+1}^k}}\norm{\param_i}{\sigma^2_{\combinedInno{k}{n+1}}} \,.
\end{align*}
Finally, the posterior predictive calculates as
\begin{align*}
 p(\statMod{n+1}^k,\combinedInno{k}{n+1}\given\stat{n}^k) = & \sum_{i=0}^1 p(\statMod{n+1}^k,\combinedInno{k}{n+1}\given\Hyp_i,\stat{n}^k)p(\Hyp_i\given\stat{n}^k)\,,
\end{align*}
with
\begin{align*}
  p(\statMod{n+1}^k,&\combinedInno{k}{n+1}\given\Hyp_i,\stat{n}^k) \\
  & = \int \norm{\tilde\param_i^k}{\sigma^2_{\statMod{n+1}^k}}\norm{\param_i}{\sigma^2_{\combinedInno{k}{n+1}}}p(\param_i\given\Hyp_i,\stat{n})\dInt\param_i\,.
\end{align*}

\bibliographystyle{IEEEtran}
\bibliography{IEEEabrv,mrabbrev,references} 

\end{document}